\documentclass[10pt,draft]{dis03}
\usepackage{epsf,amsmath}

\newcommand{\fcdz}       {\mbox{$f(c \rightarrow D^0)$}}
\newcommand{\fcdc}       {\mbox{$f(c \rightarrow D^+)$}}
\newcommand{\fcdss}      {\mbox{$f(c \rightarrow D_s^+)$}}
\newcommand{\fclc}       {\mbox{$f(c \rightarrow \Lambda_c^+)$}}
\newcommand{\fcds}       {\mbox{$f(c \rightarrow D^{\ast +})$}}
\newcommand{\dspm}       {\mbox{$D^{\ast \pm}$}}
\newcommand{\dz}         {\mbox{$D^{0}$}}
\newcommand{\dcpm}       {\mbox{$D^{\pm}$}}
\newcommand{\dsspm}      {\mbox{$D_s^{\pm}$}}
\newcommand{\lcpm}       {\mbox{$\Lambda_c^\pm$}}
\newcommand{\lcp}        {\mbox{$\Lambda_c^+$}}

\begin{document}

\title{Charm hadronisation \\
in $\gamma p$ collisions with ZEUS at HERA
}

\author{Leonid~Gladilin
\thanks{On leave from Moscow State University,
supported by the U.S.-Israel BSF} \\
(on behalf of the ZEUS Collaboration) \\
DESY, ZEUS experiment, Notkestr. 85, 22607 Hamburg, Germany \\
E-mail: gladilin@mail.desy.de}

\maketitle

\begin{abstract}
\noindent
The measured $D^{*\pm}$, $D^0$, $D^\pm$, $D_s^\pm$ and $\Lambda_c^\pm$
photoproduction cross sections
have been used to determine
charm fragmentation ratios and
fractions of c quarks hadronising as a particular charm hadron, 
$f(c \rightarrow D, \Lambda_c)$.
Events with a $\dspm$ meson produced in association with an energetic jet
have been used to measure
the charm fragmentation function.
The results are compared with different models and with previous measurements.
\end{abstract}

\section{Introduction}

Charm quark production has been extensively studied at HERA
using $\dspm$ and $\dsspm$
mesons.
The data have been compared with the theoretical predictions
by assuming the universality of charm fragmentation.
This assumption allows the charm fragmentation characteristics,
obtained in $e^+e^-$ annihilations, to be used in calculations
of charm production in $ep$ scattering.
Measuring the charm fragmentation characteristics at
HERA permits the verification of the charm-fragmentation universality
and
contributes to the knowledge of charm fragmentation.

The production of $D^{*\pm}$, $D^0$, $D^\pm$, $D_s^\pm$ and $\Lambda_c^\pm$
charm hadrons have been measured in
$e p$ scattering at HERA
in the photoproduction regime ($Q^2\approx 0$)~\cite{fratios}.
The measured production cross sections have 
been used to determine the ratio of neutral and charged $D$ meson 
production rates, $R_{u/d}$, the strangeness suppression factor, $\gamma_s$, 
the fraction of $D$ mesons produced in a vector state, $P_{\rm v}$, and
the fractions of c quarks hadronising as a particular charm hadron, 
$f(c \rightarrow D, \Lambda_c)$.
Events with a $\dspm$ meson produced in association with an energetic jet
have been used to measure the charm fragmentation function~\cite{ffunction}.

\begin{figure}[!thb]
\vspace*{7.cm}
\begin{center}
\includegraphics{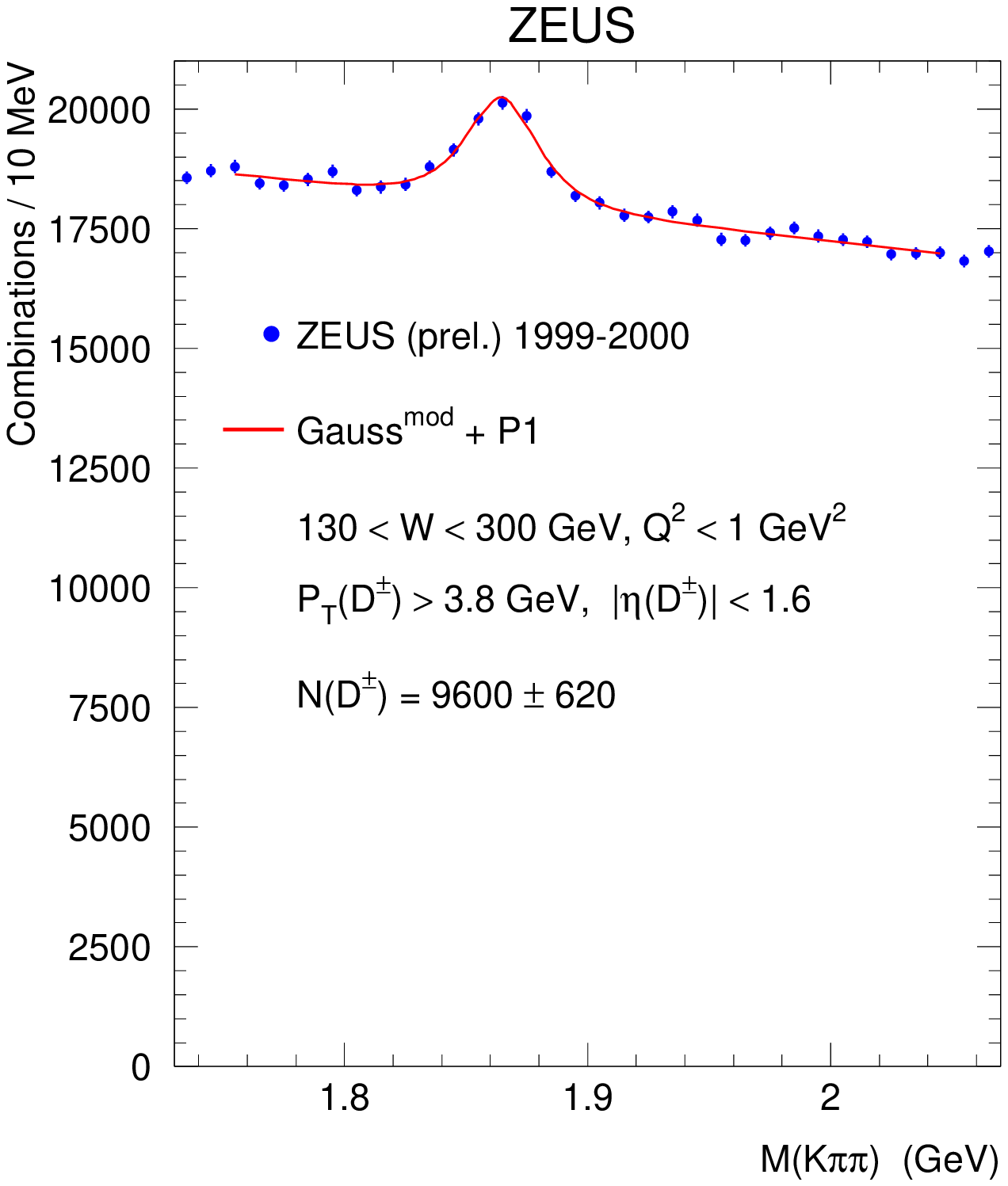}
\includegraphics{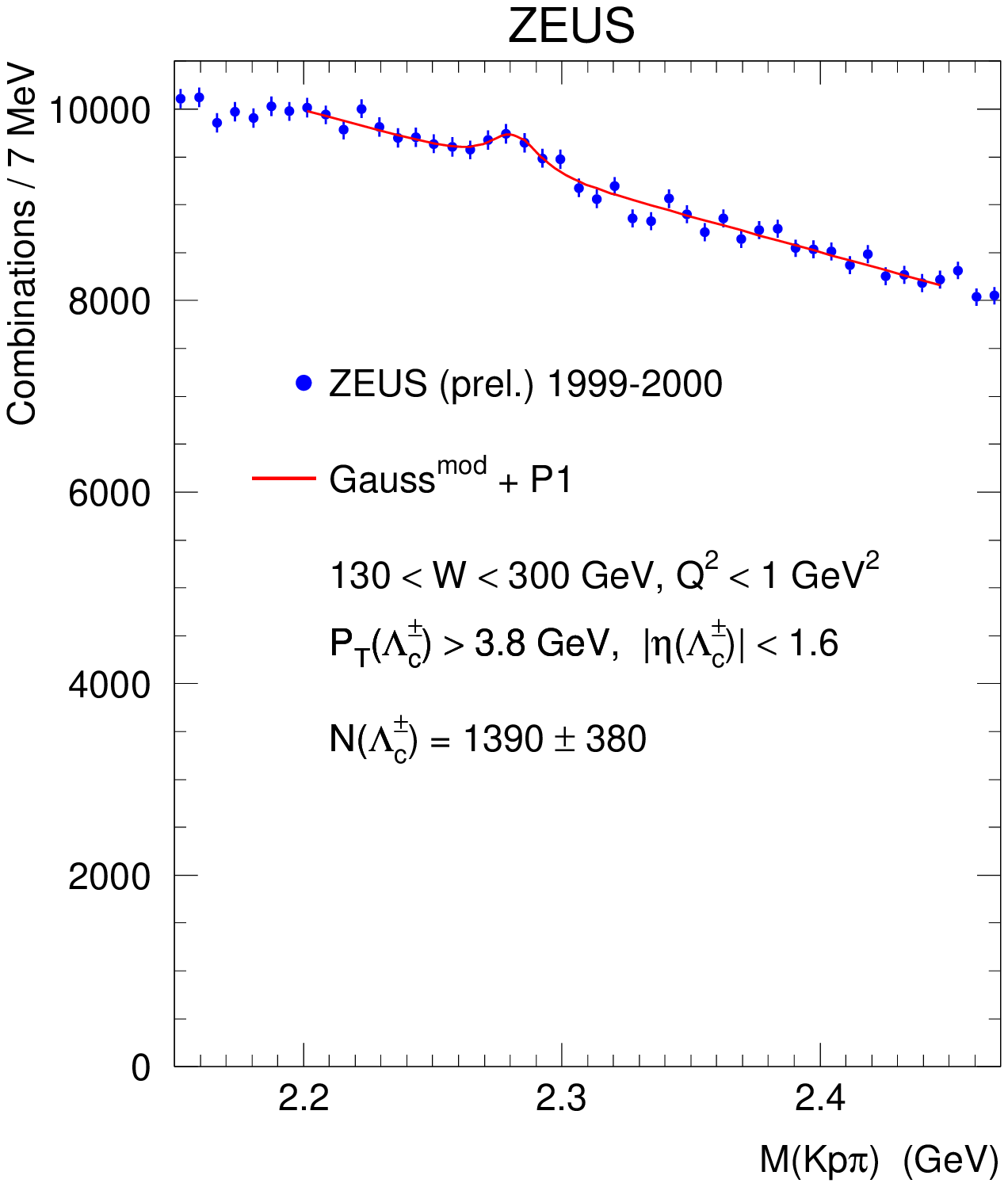}
\caption[*]{
Distributions of the reconstructed invariant mass
for the $\dcpm$ candidates (left) and for the $\lcpm$ candidates (right).
The solid curves represent fits to a sum of a modified Gaussian function
and a linear background function.
}
\end{center}
\end{figure}

\begin{table}[hbt]
\begin{center}
\begin{tabular}{|c|c|c|} \hline
& ZEUS prel. ($\gamma p$) & Combined \\
& $p_T(D,\Lambda_c)>3.8\,$GeV & $e^+e^-$ data~\cite{hep-ex-9912064} \\
& $|\eta(D,\Lambda_c)|<1.6$ &  \\
\hline
\hline
$\fcdc$ & $0.249 \pm 0.014^{+0.004}_{-0.008}$  &
$0.232 \pm 0.010$ \\
\hline
$\fcdz$ & $0.557 \pm 0.019^{+0.005}_{-0.013}$  &
$0.549 \pm 0.023$ \\
\hline
$\fcdss$ & $0.107 \pm 0.009 \pm 0.005$ &
$0.101 \pm 0.009$ \\
\hline
$\fclc$ & $0.076 \pm 0.020^{+0.017}_{-0.001}$ &
$0.076 \pm 0.007$ \\
\hline
$\fcds$ & $0.223 \pm 0.009^{+0.003}_{-0.005}$ &
$0.235 \pm 0.007$ \\
\hline
\end{tabular}
\caption{
The fractions of $c$ quarks hadronising as a particular charm hadron.
The first and second uncertainties represent statistical and systematic
uncertainties, respectively.
For the values obtained for charm production in $e^+e^-$ annihilations,
the combined statistical and systematic uncertainties are quoted.
}
\label{tab:ff}
\end{center}
\end{table}

\section{Measurement of charm fragmentation ratios and fractions}
The production of $\dspm$, $\dz$, $\dcpm$, $\dsspm$ and $\lcpm$
charm hadrons was measured in the kinematic range
$p_T(D,\Lambda_c) > 3.8\,$GeV and $|\eta(D,\Lambda_c)| < 1.6$.
The measurement was performed
for photon-proton centre-of-mass energies in the range $130<W<300\,$GeV
using an integrated luminosity of 79$\,$pb$^{-1}$.
Figure~1 shows 
distributions of the reconstructed invariant mass
for the $\dcpm$ and $\lcpm$ candidates reconstructed from the decay
channels
$D^+ \rightarrow K^-\pi^+\pi^+$ (+c.c.) and
$\lcp \rightarrow K^-p\pi^+$ (+c.c.), respectively.
The mass distributions were fitted to a sum of a ``modified''
Gaussian function and a linear background function.
The modified
Gaussian function, which was designed for the best description of
the reconstructed signals, took the form:
\begin{equation*}
{\rm Gauss}^{\rm mod}\propto \exp [-0.5 \cdot x^{1+\frac{1}{1+0.5 \cdot x}}],
\end{equation*}
where $x=|[M(K\pi)-M_0]/\sigma|$. The signal position, $M_0$,
and width, $\sigma$, as well as the numbers of charm hadrons in each signal
were free parameters of the fit.
Other details of the charm-hadron reconstruction are discussed
in~\cite{fratios}.

Using the measured cross sections, the charm fragmentation ratios are
$$R_{u/d}= 1.014 \pm 0.068 \,({\rm stat}) ^{+0.024}_{-0.031}\,({\rm syst}),$$
$$\gamma_s = 0.266 \pm 0.023 \,({\rm stat}) ^{+0.014}_{-0.012} \,({\rm syst}),$$
$$P_{\rm v} = 0.554 \pm 0.019 \,({\rm stat}) ^{+0.008}_{-0.004} \,({\rm syst}).$$

The measured $R_{u/d}$ value agrees with one. This confirms isospin
invariance which suggests $u$ and $d$ quarks are produced equally
in charm fragmentation.
The $s$ quark production is suppressed
by a factor $\approx3.5$, as the measured $\gamma_s$ value shows.
The measured $P_{\rm v}$ fraction
is sizeably smaller than the naive spin counting prediction of $0.75$.
The predictions of the thermodynamical approach~\cite{thermo}
and the string fragmentation approach~\cite{string},
which both predict $2/3$ for the fraction, are closer to,
but still above, the measured value. 

The fraction of $c$ quarks hadronising as a particular charm hadron,
$f(c\rightarrow D,\Lambda_c)$, is given by the ratio
of the production cross section for the hadron to the sum
of the production cross sections
for all charm weakly-decaying ground states.
The measured fragmentation fractions are compared in
Table~1 with the values obtained
for charm production in $e^+e^-$ annihilations~\cite{hep-ex-9912064}.
These measurements, as well as the values obtained
in deep inelastic scattering (DIS)~\cite{h1_ichep02},
agree within experimental uncertainties.
This confirms the universality of charm fragmentation.

\section{Measurement of charm fragmentation function}
Fragmentation fractions are used to parameterise the transfer
of the quark's energy to a given meson.
The measurement of the charm fragmentation function
in the transition from a charm quark to a $\dspm$ meson
was performed
for photon-proton centre-of-mass energies in the range $130<W<280\,$GeV
using an integrated luminosity of 120$\,$pb$^{-1}$.
Using events with a $\dspm$ meson produced in association with an energetic
jet, the fragmentation variable, $z$, was defined as
$$z=(E+p_{||})^{D^{*\pm}}/(E+p_{||})^{\rm jet}
\equiv (E+p_{||})^{D^{*\pm}}/2\,E^{\rm jet},$$
where $p_{||}$ is the longitudinal momentum of the $\dspm$ meson
relative to the axis of the associated jet of energy $E^{\rm jet}$.
The equivalence of $(E+p_{||})^{\rm jet}$ and $2\,E^{\rm jet}$ arises
because the jets are reconstructed as massless objects.
The $\dspm$ meson was included in the jet-finding procedure and was
thereby uniquely associated with one jet only.

\begin{figure}[!thb]
\vspace*{7.5cm}
\begin{center}
\includegraphics{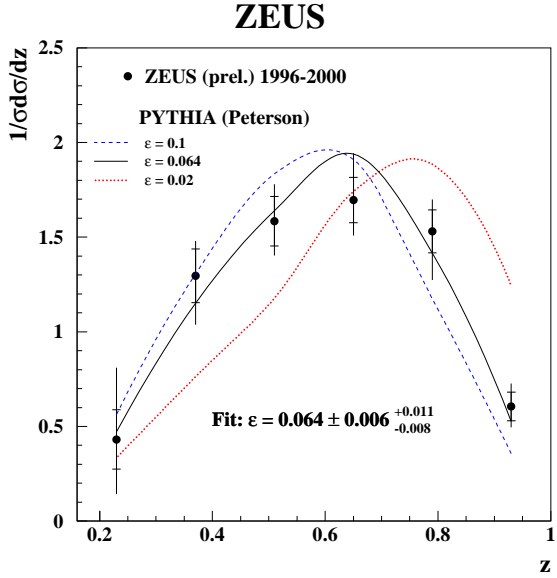}
\caption[*]{
Relative cross section $1/\sigma (d\sigma/d z)$, for the data
compared with PYTHIA predictions for different values
of the parameter $\epsilon$ in the Peterson fragmentation function.
}
\end{center}
\end{figure}

Figure~2 shows the normalised differential cross section,
$1/\sigma (d\sigma/d z)$, measured in the kinematic range
$p_T(D^{*\pm}) > 2\,$GeV, $|\eta(D^{*\pm})| < 1.5$,
$E_T^{\rm jet}>9\,$GeV and $|\eta^{\rm jet}| < 2.4$.
The data were compared with various fragmentation models
implemented in the leading-logarithmic
Monte Carlo (MC) program PYTHIA~\cite{pythia}.
The LUND string fragmentation model~\cite{lund}
modified for heavy quarks~\cite{bowler} gives a reasonable
description of the data~\cite{ffunction}.
In Fig.~2,
the measurement is compared
with PYTHIA predictions obtained using 
the Peterson fragmentation function~\cite{peterson}
with different values
of the parameter $\epsilon$.
The MC was fit to the data via a $\chi^2$-minimisation procedure
to determine the best value of $\epsilon$.
The result of the fit is
$\epsilon=0.064\pm0.006^{+0.011}_{-0.008}$.
The result is in reasonable agreement with the default value used in
PYTHIA (0.05), and with the value 0.053 obtained in the leading-logarithmic
fit~\cite{oleari}
to the ARGUS data~\cite{argus}.

\section{Summary}
The measured $D^{*\pm}$, $D^0$, $D^\pm$, $D_s^\pm$ and $\Lambda_c^\pm$
photoproduction cross sections
have been used to determine
charm fragmentation ratios and
fractions of c quarks hadronising as a particular charm hadron.
The measured ratio of neutral and charged
production rates agrees with one. This confirms isospin
invariance which suggests $u$ and $d$ quarks are produced equally
in charm fragmentation.
The $s$ quark production is suppressed
by a factor $\approx3.5$, as the measured value of the strangeness
suppression factor shows.
The measured fraction
of $D$ mesons produced in a vector state
is sizeably smaller than the naive spin counting prediction of $0.75$.

The fragmentation function for $\dspm$ mesons has been measured by requiring
a jet to be associated with the $\dspm$ meson.
The LUND string fragmentation model gives a reasonable description
of the data, as does the Peterson function with
$\epsilon=0.064\pm0.006^{+0.011}_{-0.008}$
as determined from a fit to the data.

All measured fragmentation characteristics agree with those obtained for charm
production in $e^+e^-$ annihilations, thus confirming
the universality of charm fragmentation.

\end{document}